# Hydrogen Bonds in Excited State Proton Transfer

D.A. Horke[a,b], H.M. Watts[c], A.D. Smith[c], E. Jager[c], E. Springate[d], O. Alexander[d], C. Cacho[d], R.T. Chapman[d], and R.S. Minns[c*]

[a]Center for Free-Electron Laser Science, DESY, Notkestrasse 85, 22607 Hamburg, Germany

[b]The Hamburg Centre for Ultrafast Imaging, University of Hamburg, Luruper Chaussee 149, 22761 Hamburg, Germany

[c]Chemistry, University of Southampton, Highfield, Southampton SO17 1BJ, UK

[d]Central Laser Facility, STFC Rutherford Appleton Laboratory, Didcot, Oxfordshire OX11 0QX, UK

* r.s.minns@soton.ac.uk

**Abstract**

Hydrogen bonding interactions between biological chromophores and their surrounding protein and solvent environment significantly affect the photochemical pathways of the chromophore and its biological function. A common first step in the dynamics of these systems is excited state proton transfer between the non-covalently bound molecules, which stabilises the system against dissociation and principally alters relaxation pathways. Despite such fundamental importance, studying excited state proton transfer across a hydrogen bond has proven difficult, leaving uncertainties about the mechanism. Through time-resolved photoelectron imaging measurements we demonstrate how the addition of a single hydrogen bond and the opening of an excited state proton transfer channel dramatically changes the outcome of a photochemical reaction, from rapid dissociation in the isolated chromophore, to efficient stabilisation and ground state recovery in the hydrogen bonded case, and uncover the mechanism of excited state proton transfer at a hydrogen bond, which follows sequential hydrogen and charge transfer processes.

Light driven processes in biochemistry are initiated by absorption at a central molecular chromophore which is often linked to its surrounding solvent and/or protein environment via a network of hydrogen bonds (HBs). HBs have a profound effect on the chromophore structure, and which parts of the potential energy landscape can be explored following photoexcitation. A famous example is the absence of fluorescence in the chromophore of the green fluorescent protein (GFP) when it is removed from its native protein environment.[1,2] Time- and frequency-resolved measurements on GFP have provided some remarkable results, highlighting the importance of excited state proton transfer (ESPT)[3], but the mechanisms through which this occurs, and how the protein-chromophore interactions influence the dynamics, remain unclear.[4] To circumvent the complexity of entire protein systems, bottom up approaches have centred on studies of isolated chromophores.[5-12] However, these studies do not account for the effects of the protein-chromophore interactions on the dynamics, which can dramatically alter the local electronic structure and potential energy surface. We present an alternative approach and concentrate specifically on the dynamics around a single HB in a model molecular complex. This allows us to study in detail interactions that are inaccessible in isolated chromophores and highlight the dramatic changes in dynamics and functionality caused by HBs. Using this approach we study ESPT in the ammonia dimer, a small molecular complex and model of the -NH--N- HB network present, e.g., between nucleobases in the backbone of a DNA double helix. Using time-resolved photoelectron imaging,[13,14] we uncover the two-step mechanism of ESPT across the HB and disentangle the hydrogen transfer (HT) and charge transfer (CT) processes, and how these lead to ground state recovery on an ultrafast timescale. This provides specific insight into the dynamics of the ammonia dimer but also the more general importance of ESPT and the effect that a single HB can have on the dynamics following photoexcitation.

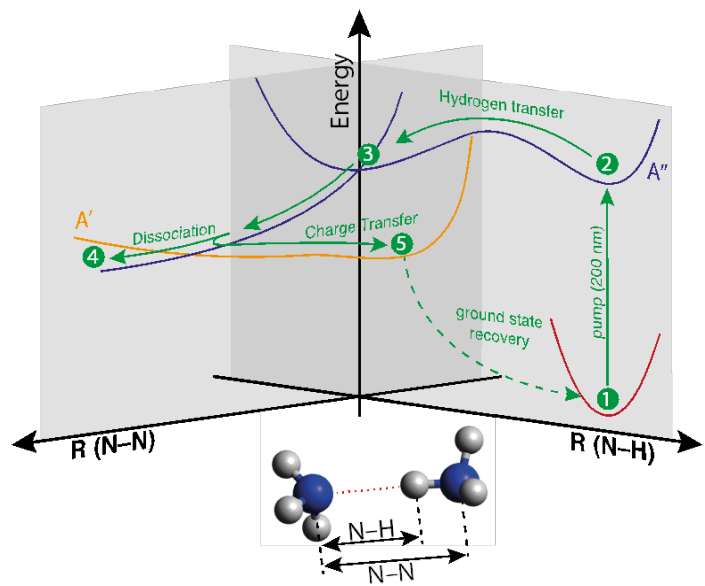

Figure 1: Schematic potential energy surfaces of the ammonia dimer along the N-N and N-H coordinates, based on calculations in [1]. Optical excitation pumps the ground state equilibrium geometry (1) into the $A''$ excited state (2), which undergoes HT (3), forming the protonated monomer in dimer structure. Dissociation on the $A''$ surface (4) leads to $NH_4$ and $NH_2$ products, or, at an extended N-N distance a conical intersection with a $A'$ state becomes accessible, leading to CT (5). The CT state can subsequently undergo internal relaxation leading to ground state recovery.

Ammonia has been the subject of numerous time- and energy-resolved experiments and theoretical calculations.[15-18] In the isolated system, UV absorption below 208 nm leads to highly efficient and ultrafast hydrogen abstraction. Studies of the photochemical dynamics of the ammonia dimer are very few by comparison. Experiments have shown that upon the formation of a molecular complex, the HB leads to an increase in the photo-excited state lifetime, reducing hydrogen abstraction to negligible levels, thereby fundamentally changing the dynamics and products formed following photoexcitation.[19-23] *Ab-initio* calculations of the excited state potential energy surface of the ammonia dimer along the HT and molecular dissociation pathways have highlighted the potential importance of HT and CT processes across the HB. However, experimental measurements have been limited to studying the initial HT process alone,[19-23] with time resolved photoelectron and photoion coincidence measurements providing a cluster-size dependent timescale of 100-330 fs.[19-23] Results of *ab-initio* calculations are shown schematically in Fig. 1, indicating the dynamical pathways accessible after photoexcitation. Following excitation into the $A''$ state (**2** in Fig. 1), a hydrogen atom is transferred across the HB to form $NH_4NH_2$, the protonated monomer in dimer (PMD) structure (**3**).[19-

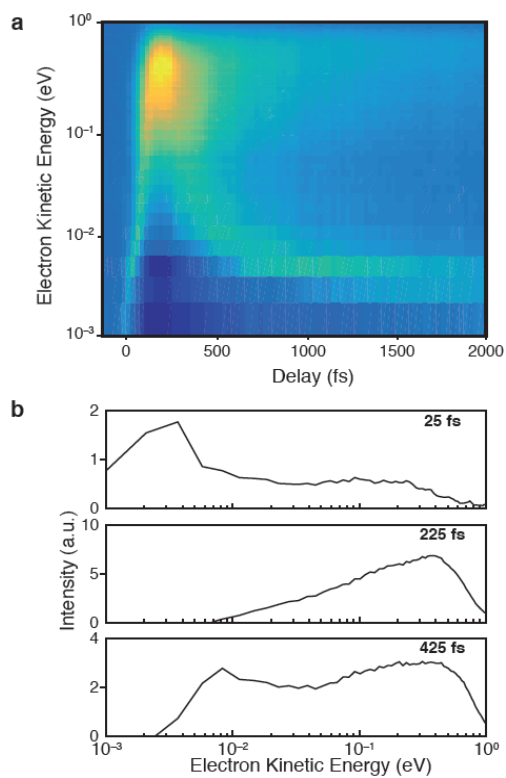

Figure 2: Time-resolved photoelectron spectra of the ammonia dimer. (a) Photoelectron intensity as a function of pump-probe delay and electron kinetic energy. (b) Representative photoelectron spectra at three pump-probe delays, directly after optical excitation (top), following population of the PMD structure (center), and after the wavepacket motion (bottom).

23] This stabilises the complex and prevents hydrogen abstraction, in stark contrast to the dynamics observed in the monomer. The dynamics following population of the PMD structure have not been observed experimentally or calculated theoretically. The available static potential energy surfaces suggest that two competing decay pathways are accessible; dissociation on the $A''$ state surface with loss of a $NH_2$ fragment (**4**) and CT via a conical intersection with the $A'$ state (**5**), eventually followed by reverse proton transfer and ground state recovery. The dynamics following population of the PMD structure crucially determine the fate of the complex, i.e. dissociation or ground state recovery, yet neither experimental not theoretical approaches have so far provided an insight into these dynamics, and how they relate to ESPT processes in hydrogen bonded systems in general.

Results and discussion

Time-resolved photoelectron spectra, following excitation of the dimer into the $A''$ state at 200 nm (6.1 eV) and probing the system through photoionization at 395 nm (3.1 eV), are presented in Fig. 2a. In Fig. 2b we highlight the time dependence by plotting the total photoelectron signal at three delays. At the time of photoexcitation (i.e., $t = 0$ fs) we observe emission at very low electron kinetic energies (eKE) (< 0.1 eV), consistent with the ionization potential of the ammonia dimer of 9.19 eV,[20,24,25] considering the total photon energy deposited into the system is ~9.3 eV. The comparatively low signal levels observed are due to poor Franck-Condon overlap of the dimer equilibrium geometry with the corresponding lowest ion state. Over the next few hundred femtoseconds the photoelectron signal shifts to higher eKEs and grows in intensity, directly monitoring the motion along the potential energy surface of the *A″* state, into the protonated monomer in dimer (PMD) structure (**3** in Fig. 1). The higher observed eKEs are due to a reduction in ionization potential in the PMD structure to 8.62 eV,[20] yielding eKEs up to ~0.7 eV, Fig. 2b. This spectral shift maps the wavepacket motion along the *A″* potential energy surface into the PMD structure (**3** in Fig. 1). We quantify this by fitting spectral slices through the data with a biexponential decay model (see ESI for details [26]). While the decay constants extracted from these fits for the low energy slices are not representative of the dynamics, due to the spectral shifting of intensity leading to repopulation of lower energy bands at later times, the onset of population arriving in a spectral band, i.e. the appearance delay, is nonetheless accurate. In Fig. 3a we plot this delay as a function of eKE. Initially, a clear linear dependence is observed, which flattens out in the spectral region 0.5-0.6 eV, as the curvature of the potential energy surface is reduced and the local potential minimum of the PMD structure is reached. We note that the temporal shift is independent of the photoelectron signal levels, and the slowing down of the wavepacket motion cannot be explained by a loss of Franck-Condon overlap. The shifting photoelectron intensity therefore provides a timescale of 130 fs for the HT process and motion into the PMD geometry, in good agreement with previous measurements.[20,22,23]

The signal observed following population of the PMD structure, an intense broad peak with eKEs from 0.1 to 0.7 eV, shows a biexponential decay. Figure 3b shows the integrated signal for spectral slices through the eKE distributions as a function of time. The fit parameters change as a function of eKE, with the shift in energy causing each band to have different associated lifetimes and amplitudes (see ESI for details). We note that at very low eKE (< 0.1 eV) the data cannot be easily fit with a simple kinetic model due to wavepacket dynamics leading to population transfer in and out of this band, and hence a complex photoelectron intensity distribution with multiple peaks. We attribute the biexponential decay of the photoelectron signal to a bifurcation of the wavepacket in the PMD geometry, where two relaxation pathways are accessible. The short lifetime decay component is a consequence of the shift of photoelectron intensity to lower eKE. The shift again involves the motion of the wavepacket along the potential energy landscape leading to a change in effective ionization potential, as can clearly be seen from the time- and energy-resolved photoelectron intensity map in Fig. 2a, and the reappearance of a peak at low eKE, Fig. 2b. In the temporal picture, Fig. 3b, the repopulation of the low eKE band leads to the emergence of a second peak in the 0.01 – 0.1 eV spectral region after 350 fs.

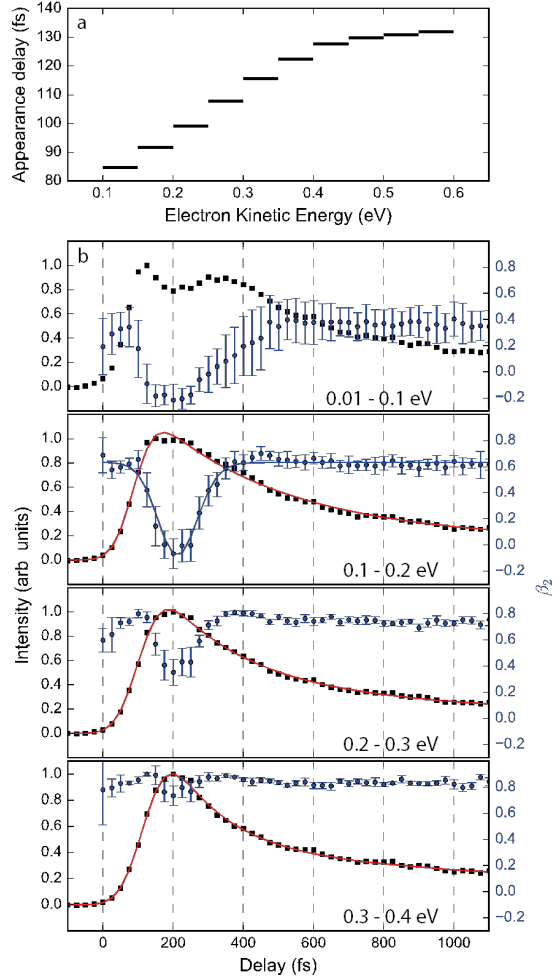

Figure 3: (a) Appearance delay of photoelectron features as a function of electron kinetic energy, mapping out the motion along the $A''$ surface. (b) Photoelectron intensity integrated over spectral slices as a function of pump-probe delay (black squares), and corresponding $\beta_2$ anisotropy parameter (blue circles). Shown in red are biexponential fits to the measured intensity distributions (see text and ESI for details.).

To identify the origin of the spectral shifting and biexponential decay, as well as the nature of the involved electronic states, we turn to the photoelectron angular distributions (PADs). These are analysed taking into account $\beta_2$ and $\beta_4$ anisotropy parameters,[27] (see ESI for details[26]). Extracted $\beta_2$ parameters are shown in Fig. 4 as a function of eKE and pump-probe delay. Shown in blue in Fig. 3b are the average anisotropy parameter for energetic slices through the photoelectron distributions. For the majority of the time- and energy-space investigated we observe strongly positive $\beta_2$ parameters. We rationalize these by considering the symmetry of the involved electronic states, and hence the expected outgoing photoelectron partial waves. In ammonia, UV excitation leads to promotion of an electron from the lone pair orbital into a nitrogen-centred Rydberg orbital of predominantly 3*s* character (in the Franck-Condon region), which subsequently evolves into $\sigma^*$ character at extended N-H bond lengths.[22] The transition in the dimer is comparable to the monomer transition, but correlates with the NH$_4$(3*s*)—NH$_2$(X) electronic structure formed following HT.[18] Photoionisation from the excited state therefore proceeds from an orbital of predominantly 3*s* Rydberg character, and results in the emission of photoelectrons with *p*-type partial waves. In the laboratory frame this is observable as PADs that peak along the laser polarization axis, resulting in positive $\beta_2$. This is consistent with our observations at early delay times, i.e. directly after photoexcitation. Strongly positive $\beta_2$ are also observed at long delay times ($t > 500$ fs). The anisotropy parameters are quantitatively identical to those seen at early times indicating that the long-time dynamics are also occurring on the initially excited $A''$ state. As the total excitation energy (6.2 eV) is above the adiabatic dissociation threshold of the $A''$ state (4.9 eV[20]), we assign the long-time decay component, with a typical lifetime of ~2 ps (see ESI for details[26]), to dissociation of the dimer into $NH_4$ and $NH_2$ fragments from the PMD structure. The comparatively long lifetime of dissociation into $NH_4$ and $NH_2$ fragments is due to the rather flat potential along the N-N reaction coordinate. This provides little driving force towards the dissociation process, extending the lifetime and allowing other decay processes to compete and dominate the dynamics of the complex. This decay channel, while peaked at 0.5 eV eKE, leads to the emission of photoelectrons with a large energy bandwidth, as seen in Fig. 2b, due to the shallow potential minimum of the PMD.

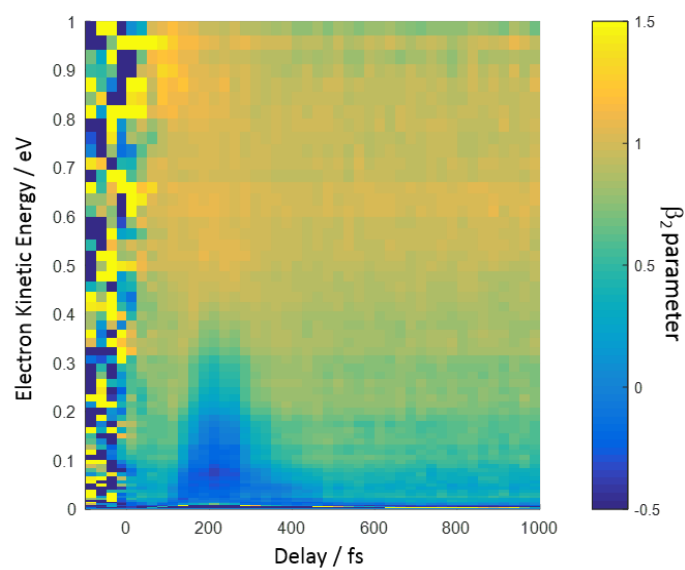

Figure 4: Time-resolved Photoelectron anisotropy. The transient reduction observed around 200 fs is indicative of a non-adiabatic electronic transition into an excited state of different symmetry.

At intermediate delay times, in the region between ~100 and 500 fs, we observe a strong transient reduction in $\beta_2$, which exhibits a temporal and spectral dependence, Fig. 4. The observed dependence on eKE correlates with the disappearance of the short lifetime decay component, as can be seen in Fig. 3b, where the reduction in $\beta_2$ becomes significantly more pronounced towards lower energies. We also note a temporal offset between the peak of the photoelectron intensity and the minimum $\beta_2$ parameter of around 50 fs, clearly identifiable in the spectral regions 0.01 – 0.1 eV and 0.1 – 0.2 eV. The transient reduction in photoelectron anisotropy suggests that this effect is not due to changes in nuclear geometry (i.e. wavepacket motion) or the different scattering processes this may drive, but due to an electronic state transition[28] leading to a loss of Rydberg character.

The correlation of the reduction in $\beta_2$ with the shifting photoelectron intensity, combined with the temporal offset, points to a 2-step decay mechanism. This first step involves motion along the $A''$ electronic state surface away from the PMD structure. After ~50 fs the wavepacket reaches a crossing point where a non-adiabatic electronic transition into a state of different symmetry occurs, leading to the observed changes in the PADs. Based on the potentials of Farmanara *et al*[1], the non-adiabatic transition leads to population of the $A'$ state, which has a higher binding energy and CT character, making it less Rydberg like and resulting in a reduction of the measured $\beta_2$ parameter. The wavepacket then continues on the newly populated state towards geometries with higher effective ionisation potentials, reducing the observed eKE. The measured changes in eKE and $\beta_2$ parameter support this assignment of the non-adiabatic transition into the $A'$ state. The wavepacket continues on its path towards the ground state until it leaves our observation window, defined by the probe photon energy, in ~ 400 fs.

The observations can be rationalised in terms of the dynamics outlined in figure 1: Following formation of the PMD structure the molecule moves towards longer N-N bond distances. After approximately 50 fs the complex reaches an N-N separation of ~ 4Å where the $A''$ state crosses another excited state of $A'$ symmetry at a conical intersection.[20] Access to the conical intersection is not significantly delayed by the relatively shallow potential along the N-N coordinate due to its proximity to the Franck-Condon geometry. The Franck-Condon geometry has an N-N bond length of ~3.6 Å with a broad distribution due to the shallow potential of the ground electronic state. The conical intersection is therefore readily accessible following HT. Motion of the wavepacket through the conical intersection leads to a non-adiabatic transition into the $A'$ state, undergoing CT to form the energetically more stable $NH_4^+NH_2^-$ structure. The change in electronic structure leads to the observed change in $\beta_2$ parameter. The shape of the $A'$ potential energy surface drives the wavepacket towards shorter N-N bond lengths, reducing the barrier for reverse proton transfer. At N-N bond lengths below 2.8 Å the barrier to reverse proton transfer vanishes, such that the PMD geometry is unstable and forms a saddle point on the $A'$ excited state potential energy surface.[20] The $A'$ state therefore provides a barrierless pathway back to the electronic ground state through a reverse proton transfer process.[20] The relaxation process leads to an increase in the effective ionisation potential with the wavepacket moving out of our observation window in ~ 400 fs. Finally, following repopulation of the electronic ground state, the excess vibrational energy of the dimer is far above the dimer binding energy (0.16 eV[29]) and cannot be dissipated to the surroundings such that the complex will dissociate into two ammonia molecules.

We can therefore summarise the full dynamics of the system as follows; photoexcitation of the ammonia dimer populates the $A''$ electronic excited state. This undergoes rapid HT in ~ 130 fs, forming the PMD structure. Following HT, two decay pathways are observed and the wavepacket bifurcates. Part of the population undergoes a non-adiabatic transition into the $A'$ through a conical intersection in ~ 50 fs, leading to CT and the formation of the $NH_4^+NH_2^-$ structure. This population in the CT state subsequently relaxes through reverse proton transfer reforming the electronic ground state of the dimer, which dissociates into two $NH_3$ fragments. The remaining population in the $A''$ state dissociates into $NH_2$ and $NH_4$ products with a typical lifetime of 2 ps.

The differences between the dynamics of the complex with the isolated molecule following excitation into the $A''$ state are striking. Where isolated ammonia exhibits a high probability for dissociation through hydrogen abstraction, the ammonia dimer shows an efficient pathway back to the electronic ground state, enabled by a single hydrogen bonding interaction. Considering the binding energy of the HB is only 3% of the excess energy deposited into the system by the pump photon (0.16 eV binding energy compared with the 6.1 eV photon energy), it is surprising that it should cause such a dramatic change in the dynamics. The hydrogen-bonding interaction acts to effectively stabilise the molecule against hydrogen abstraction, providing an efficient pathway for population to return to the ground state. This is enabled by sequential hydrogen and CT processes, essentially an excited-state proton transfer, suggesting that this could also be a sequential two-step process in other biological system, such as photoactive proteins.[3] In condensed phase or *in vivo* biological systems, containing large hydrogen bonding networks, efficient dissipation of the excess excitation energy into the surroundings will stabilise the recovered ground state system against dissociation, potentially explaining the ubiquity of HBs in biological systems, and the crucial role they play in photostability. Our contribution furthermore highlights the importance of relatively weak perturbations in defining excited state dynamics. The presence of a single HB influences not only the structure of a system and introduces additional steric constraints, but additionally leads to significant changes to the electronic structure of excited states and which parts of the potential energy landscape are accessible following photoexcitation.

**Acknowledgements**

All authors thank the STFC for access to the Artemis facility (Grant No. 13220015). R.S.M. thanks the Royal Society for a University Research Fellowship (Grant No. UF100047) and the Leverhulme Trust for research support and for the studentship of A. D. S. (Grant No. RPG-2013-365). H.M.W. thanks the Central Laser Facility and Chemistry at the University of Southampton for a studentship. E. J. thanks Chemistry at the University of Southampton for a studentship. We also acknowledge funding from the EC's Seventh Framework Programme (LASERLAB-EUROPE, Grant No. 228334). This work has been supported by the excellence cluster "The Hamburg Center for Ultrafast Imaging-Structure, Dynamics and Control of Matter at the Atomic Scale" of the Deutsche Forschungsgemeinschaft (CUI, DFG-EXC1074). D.A.H. was supported by the European Research Council through the Consolidator Grant Küpper-614507-COMOTION. We thank Phil Rice for technical assistance.

**References**

[1]  M. W. Forbes and R. A. Jockusch, J. Am. Chem. Soc. **131**, 17038 (2009).
[2]  H. Niwa, S. Inouye, T. Hirano, T. Matsuno, S. Kojima, M. Kubota, M. Ohashi, and F. I. Tsuji, Proc. Natl. Acad. Sci. U. S. A. **93**, 13617 (1996).
[3]  C. Fang, R. R. Frontiera, R. Tran, and R. A. Mathies, Nature **462**, 200 (2009).
[4]  J. J. van Thor, Chem. Soc. Rev. **38**, 2935 (2009).
[5]  F. Buchner, A. Nakayama, S. Yamazaki, H. H. Ritze, and A. Lubcke, J. Am. Chem. Soc. **137**, 2931 (2015).
[6]  A. S. Chatterley, C. W. West, G. M. Roberts, V. G. Stavros, and J. R. R. Verlet, J. Phys. Chem. Lett. **5**, 843 (2014).
[7]  C. E. Crespo-Hernandez, B. Cohen, P. M. Hare, and B. Kohler, Chem. Rev. **104**, 1977 (2004).
[8]  J. B. Greenwood, J. Miles, S. De Camillis, P. Mulholland, L. J. Zhang, M. A. Parkes, H. C. Hailes, and H. H. Fielding, J. Phys. Chem. Lett. **5**, 3588 (2014).
[9]  D. A. Horke and J. R. R. Verlet, Phys. Chem. Chem. Phys. **14**, 8511 (2012).
[10]  C. R. S. Mooney, D. A. Horke, A. S. Chatterley, A. Simperler, H. H. Fielding, and J. R. R. Verlet, Chem. Sci. **4**, 921 (2013).
[11]  N. M. Webber and S. R. Meech, Photochem. Photobiol. Sci. **6**, 976 (2007).
[12]  C. W. West, J. N. Bull, A. S. Hudson, S. L. Cobb, and J. R. R. Verlet, J. Phys. Chem. B **119**, 3982 (2015).
[13]  G. R. Wu, P. Hockett, and A. Stolow, Phys. Chem. Chem. Phys. **13**, 18447 (2011).
[14]  K. L. Reid, Mol. Phys. **110**, 131 (2012).
[15]  J. Biesner, L. Schnieder, G. Ahlers, X. X. Xie, K. H. Welge, M. N. R. Ashfold, and R. N. Dixon, J. Chem. Phys. **91**, 2901 (1989).
[16]  A. S. Chatterley, G. M. Roberts, and V. G. Stavros, The Journal of Chemical Physics **139**, 034318 (2013).
[17]  V. M. Donnelly, A. P. Baronavski, and J. R. McDonald, Chem. Phys. **43**, 271 (1979).
[18]  H. Yu, N. L. Evans, A. S. Chatterley, G. M. Roberts, V. G. Stayros, and S. Ullrich, J. Phys. Chem. A **118**, 9438 (2014).
[19]  F. Farmanara, H. H. Ritze, V. Stert, W. Radloff, and I. V. Hertel, Eur. Phys. J. D **19**, 193 (2002).
[20]  P. Farmanara, W. Radloff, V. Stert, H. H. Ritze, and I. V. Hertel, J. Chem. Phys. **111**, 633 (1999).
[21]  P. Farmanara, V. Stert, H. H. Ritze, W. Radloff, and I. V. Hertel, J. Chem. Phys. **115**, 277 (2001).
[22]  R. Farmanara, H. H. Ritze, V. Stert, W. Radloff, and I. V. Hertel, J. Chem. Phys. **116**, 1443 (2002).
[23]  V. Stert, W. Radloff, C. P. Schulz, and I. V. Hertel, Eur. Phys. J. D **5**, 97 (1999).
[24]  F. Carnovale, J. B. Peel, and R. G. Rothwell, J. Chem. Phys. **85**, 6261 (1986).
[25]  A. Lindblad, H. Bergersen, W. Pokapanich, M. Tchaplyguine, G. Ohrwall, and O. Bjorneholm, Phys. Chem. Chem. Phys. **11**, 1758 (2009).
[26]  See supplementary material [URL] which contains references [30-33].
[27]  J. Cooper and R. N. Zare, The Journal of Chemical Physics **48**, 942 (1968).

[28]     J. Lecointre, G. M. Roberts, D. A. Horke, and J. R. R. Verlet, J. Phys. Chem. A **114**, 11216 (2010).
[29]     A. S. Case, C. G. Heid, S. H. Kable, and F. F. Crim, J. Chem. Phys. **135**, 084312 (2011).
[30]     A. Eppink and D. H. Parker, Rev. Sci. Instrum. 68, 3477 (1997).
[31]     D. Irimia, D. Dobrikov, R. Kortekaas, H. Voet, D. A. van den Ende, W. A. Groen, and M. H. M. Janssen, Rev. Sci. Instrum. 80, 113303 (2009).
[32]     G. M. Roberts, J. L. Nixon, J. Lecointre, E. Wrede, and J. R. R. Verlet, Rev. Sci. Instrum. 80, 053104 (2009).
[33]     F. Frassetto, C. Cacho, C. A. Froud, I. C. E. Turcu, P. Villoresi, W. A. Bryan, E. Springate, and L. Poletto, Opt. Express 19, 19169 (2011).